\documentclass[aps,amsmath,twocolumn]{revtex4}
\usepackage{graphicx}
\usepackage{relsize}
\usepackage{enumitem}
\setlist{nolistsep,leftmargin=*} %compress space for all lists

\begin{document}

%\voffset=0.75truein

\title{The search for life and a new logic}

\author{Douglas Scott} \email{docslugtoast@phas.ubc.ca}
\author{Ali Frolop} \email{afrolop@phas.ubc.ca}
\affiliation{Dept.\ of Physics \& Astronomy,
 University of British Columbia, Vancouver, Canada}

\date{1st April 2020}

\begin{abstract}
Exploring the Universe is one of the great unifying themes of humanity.
Part of this endeavour is the search for extraterrestrial life.
But how likely is it that we will find life, or that if we do
it will be similar
to ourselves?  And therefore how do we know where and how to look?
We give examples of the sort of reasoning that has been used to narrow and
focus this search and we argue that obvious extensions to that logical
framework will result in greater success.
\end{abstract}

\maketitle

\noindent
\section{Introduction}
Motivations given for Solar System exploration missions, as well as for
studies of exoplanets,
often have the search for life at the very top of the list.  Picking some
examples,
the stated science goals for the whole of NASA's Mars Exploration Program
are to ``study Mars as a planetary system in order to understand the formation
and early evolution of Mars as a planet, the history of geological processes
that have shaped Mars through time, {\it the potential for Mars to have hosted
life}, and the future exploration of Mars by humans'',\cite{MEP} while
in Europe ``The goals of ExoMars are to search for {\it signs of past life on
Mars\/}''.\cite{ExoMars}  Elsewhere in the Solar System, the aims of the
Dragonfly mission to Titan are
``to search for chemical signatures that could indicate
{\it water-based and/or hydrocarbon-based life\/}''\cite{Dragonfly} and
the Europa Clipper will
``investigate whether the icy moon {\it could harbor conditions
suitable for life\/}''.\cite{EuropaClipper}  Moving further afield,
``The Origins Space Telescope will trace the history of our origins from the
time dust and heavy elements permanently altered the cosmic landscape {\it to
present-day life} \dots {\it How common are life-bearing
worlds?\/}''\cite{Origins} and
``The Habitable Exoplanet Observatory is a concept for a mission to
\dots {\it search for signatures of habitability such as water},
and be sensitive to gases in the atmosphere possibly {\it indicative of
biological activity}, such as oxygen or ozone''.\cite{HabEx}

Beyond these few examples, there are countless others.
In general, astronomers cannot talk about planetary exploration or
exoplanetary observational studies for more than a couple of sentences
without mentioning the search for life.

Is this reasonable?  Is there no motivation for studying a planet other than
to search for life?  While some cynical people might suggest that the reasons
for this single-minded focus are sociological or political \cite{OrFinancial},
we are merely scientists, and so in this paper we will concentrate only on
what rational thinking can say about this question.  Let us turn to the most
basic aspect of the scientific process, namely logic.  There is a famous
syllogism that illustrates how logical reasoning works:
\vspace{0.2cm}
\begin{itemize}
\item{\sf A}\ All elephants are grey.
\item{\sf B} Mice are grey.
\item{\sf C} Therefore mice are elephants. \cite{grey}
\end{itemize}
\vspace{0.2cm}

The search for life elsewhere in the Universe follows a similar form of
dialectical thinking:
\vspace{0.2cm}
\begin{itemize}
\item{\sf A} The Earth has life.
\item{\sf B} Some other places are like the Earth.
\item{\sf C} Therefore these other places have life.
\end{itemize}
\vspace{0.2cm}

We will suggest that this is not only logically sound, but that extending
such reasoning gives us a way to select specific places where life is much
more likely to be found, as we will discuss in Section~V.

\section{Historical digression}
First, let us go back to the time of ancient Greece \cite{BigBangTheory}, when
several philosophers, notably Leucippus \cite{existed},
Democritus and Epicurus, argued that
the Universe was large and contained a multitude of life-bearing worlds.
This idea of
``Cosmic pluralism'' was continued by Middle Eastern scholars and
was promoted in Europe by Giordano Bruno, among others.
It was formalised in the 1686 book
``Entretiens sur la pluralit{\'e} des mondes'' \cite{Fontenelle}
by Bernard Le Bovier de Fontenelle.  Deeply intertwined with
religious thinking \cite{religion},
the basic concept was that the Creator would surely not
have made all these worlds without purpose, and hence each world must have
been made for its inhabitants.

As Sir David Brewster \cite{kaleidoscope} put it, when
``we trace throughout all the heavenly bodies the same
uniformity of plan, is it possible to resist the influence that there is
likewise an uniformity of
purpose; so that if we find a number of spheres linked together by the same
bond, and governed by the same laws of matter, we are entitled to conclude that
the end for which one of these was constituted, must be the great general end
of all, -- to become a home of rational and God-glorifying creatures''.
To rephrase this argument:
\vspace{0.2cm}
\begin{itemize}
\item{\sf A} The Earth was made for humans.
\item{\sf B} Other planets exist.
\item{\sf C} Therefore there are beings on all other planets.
\end{itemize}
\vspace{0.2cm}

This ``plurality of worlds'' and cosmic-abundance-of-life concept
was popular in the 17th, 18th and 19th centuries.  It was promoted by such
luminaries as Adams, Herschel, Huygens, Locke, and Newton.  Camille
Flammarion's 1862 book specifically devoted to the topic,
``La Pluralit{\'e} des mondes habit{\'e}s'' \cite{Flammarion},
went through 33 editions in 21 years
and includes statements such as ``we who inhabit this world are only a few out
of all the worlds''.

In 1837, popular astronomy author Reverend Thomas Dick \cite{dick} went through
a series of five arguments for life on other worlds in the Solar System, leading
to an estimate that there were 21,894,974,404,480 inhabitants
in total \cite{factors}; he did not include the Sun
in his calculation, although he acknowledged that its surface area would
allow for a larger number of inhabitants than all of the planetary bodies.
However, William Herschel
had already stated that ``we need not hesitate to admit
that the sun is richly stored with inhabitants''.\cite{WHerschel}
Moreover, astronomer Johann Bode \cite{BodesLaw},
describing the inhabitants of the Sun, stated:
``Who would doubt their existence? The most wise author of the world assigns
an insect lodging on a grain of sand and will certainly not permit \dots
the great ball of the sun to be empty of creatures and still less of rational
inhabitants who are ready gratefully to praise the author of life''.

Herschel also talked about the Moon, stating in 1780 that there was a ``great
probability, not to say almost absolute certainty, of her being
inhabited''\cite{HMoon1} and in 1795 he added that ``the analogies that have
been mentioned are fully sufficient to establish the high probability of
the moon's being inhabited like the Earth''.\cite{HMoon2}
It had already been known since the time of Galileo, that the Moon
possessed seas and volcanic craters.
However, further evidence of life appeared in a series of articles published
in The Sun newspaper in New York in 1835 \cite{hoax}, based on
new observations by William Herschel's son John.  These articles
discuss how forests, fields and beaches could be seen on the lunar surface,
and with a little more scrutiny, bisons and sheep,
as well as bipedal beavers, blue goats, unicorns and man-bats.\cite{manbats}

Hence we see that, during the 19th century, the Solar System was understood
to be teeming with a great variety of living creatures, and presumably the
rest of the Universe also.  Following the usual logic,
William Herschel's final conclusion was that
``if stars are suns, and suns are inhabitable, we see at once what an
extensive field for animation opens itself to our view''.\cite{HStars}

\section{Mars}

Proponents of the study of the biota of Mars are in good company,
since they are following
the same lines of reasoning as the champions of ``cosmic plurality'',
namely
\vspace{0.2cm}
\begin{itemize}
\item{\sf A} Earth has life.
\item{\sf B} Mars is similar to Earth.
\item{\sf C} Therefore Mars has (or did have) life.
\end{itemize}
\vspace{0.2cm}

Since the 17th century, we have known that the rotational period of
Mars is approximately the same as the Earth's.  Over time, improvements
in the measurements grew along with the ideas of ``cosmic plurality''.
Hence, as it became clearer that a Mars day is very similar to an Earth day,
there was growing obsession with the
question: is there Life on Mars?\cite{LifeOnMars}
This quest was also encouraged by apparent evidence for water on the planet,
including the famous canals \cite{canali} seen by Percival Lowell.\cite{Lowell}
Thus followed decades of Martians appearing in books, motion pictures and radio
broadcasts.\cite{Welles}

More recently,
many missions to Mars have focused on the search for evidence of
biological activity.  Although there are continuing claims that such
evidence has been found, the general consensus is that Mars might
be barren today.  But since Mars is so similar to Earth, and the logic is 
so unassailable, then if Mars has no life now, it must be that it
had life at some other time.  Hence attention has focused on looking for
evidence of water on ancient Mars.

\section{Water}

Our home planet is about 70\,\% covered with water and swarming with
living organisms (if not necessarily intelligent living organisms
\cite{GalaxySong}).  By the now-familiar logic, it is obvious
that liquid water is necessary for the development and
sustainment of life.  In other words:
\vspace{0.2cm}
\begin{itemize}
\item{\sf A} The Earth has water.
\item{\sf B} The Earth has life.
\item{\sf C} Therefore, where there is water there must be life.
\end{itemize}
\vspace{0.2cm}

``Habitability'' then equates to the presence of ${\rm H}_2{\rm O}$,
not as ice or steam, but in its liquid form.\cite{water}
A planet in a habitable region is also referred to as being in the
``Goldilocks Zone''.\cite{porridge}

But how do we know we are looking in the right places for life?
We simply defer to the so-called
``streetlight effect'', which states that usually the light has been placed
in just the right place for you to be able to see the thing you are
looking for.  This follows the same exacting rules of deduction that we have
described above.

\section{The new logic}

To further extend this line of reasoning, might we not expect that bodies
sharing further attributes with the Earth will have a higher chance of
harbouring life?

Our planet has several characteristics that make it special.
For example, Earth's orbital inclination is very close to zero \cite{zero}
-- hence we should look
for life on planets with almost no orbital inclination.
Perhaps we should always focus our attention on the third planet from
the star in any exoplanet system, or on the fifth largest planet?

Earth is also the greenest place in the Solar System, suggesting that we
should search for life on planets with the same colour as
the Earth.\cite{plants}
Additionally we only have a single, large moon, which may be beneficial
for life, \cite{asimov} and hence we can ignore planets with too few or
too many moons.

A particularly fruitful search may be in any planetary systems we find
that initially look like they have nine planets but turn out to have only
eight.

However, we have only been considering the obvious reasons that the Earth
is special.  Following the thinking described earlier in this paper, it is
clear that {\it any\/} characteristic similar to Earth's should
make life compulsory, according to pure logic.  Hence other bodies whose
names also
start with the letter ``E'' should be good bets.  In fact this has already been
confirmed in the Solar System, where Europa and Enceladus have been
highlighted for future searches for life.

Another popular place to look is Titan, and, while it does not start with an
``E'', it has the same number of letters as ``Earth'', making it another
obvious target.  Moreover, it starts with the same letter as ``Terra'',
the Latin name for Earth.\cite{Titan}

Maybe we should concentrate on places with lots of Na$\,$Cl \cite{salt},
while avoiding those with almost none? \cite{nothing}
As a last suggestion, perhaps planets whose names mean ``dirt'' in one
of their native languages are likely to host life? \cite{dirt}

We hope that some of these ideas \cite{us} will be pursued by future 
targeted exoplanet observations, as well as SETI searches.  Following the
same rigorous logic that has been applied by centuries of researchers of
extraterrestrial life, we hope that readers of this paper will come up
with visionary ideas of their own.\cite{further}

%%%%%%%%%%%%%%%%%%%%%%%%%%%%%%%%%%%%%%%%%%%%%%%%%%%%%%%%%%%%%%%%%
%%%
%%%                     BIBLIOGRAPHY
%%%
%%%%%%%%%%%%%%%%%%%%%%%%%%%%%%%%%%%%%%%%%%%%%%%%%%%%%%%%%%%%%%%%%

\smallskip

%\newpage
%\vskip .75 in


\begin{thebibliography}{9}
\bibitem{grey}We shouldn't be distracted by the fact that neither elephants nor
mice are actually always grey.  Notwithstanding, the logic is irrefutable.
\bibitem{BigBangTheory} A common way to start explaining physics:
``It's a warm summer evening in ancient Greece \dots'', Cooper S., in
``Big Bang Theory'', Ser.~3, Ep.~10.
\bibitem{existed} Often called the Father of Atomism.  If he existed, that is.
\bibitem{religion} Normally we try to avoid discussing religion in our papers,
but, Lord knows, it's sometimes hard to avoid.
\bibitem{factors} This number, nearly 22 trillion, has as its prime factors
2, 5, 19, 10067 and 178859.  Dr.\ Frolop is studying the significance of this.
\bibitem{hoax} Reprinted from the Edinburgh Journal of Science.  It may
be significant that these discussions of the abundance of life appeared in
``The Sun''.
\bibitem{canali} The linear features were first spotted by the astronomer
Giovanni Schiaparelli and the fever was stoked by a mistranslation of his
use of the Italian word ``cannelloni''.
\bibitem{porridge} Perhaps the name is because the porridge in some versions of
the story is very watery.
\bibitem{zero} Some people claim that it's exactly zero.
\bibitem{nothing} Since they're like nothing on Earth.
\end{thebibliography}
\end{document}